\documentclass[useAMS]{mn2e}
\usepackage{lscape}
\usepackage{rotating}
\usepackage{amssymb}
\usepackage{float}

\def\av{\hbox{A$_{\rm V}$}}

\def\msun{\hbox{M$_\odot$}}

\def\t4{\hbox{t$_{\rm 4}$}}

\def\cm3{\hbox{cm$^{-3}$}}

\input psfig.sty
\input epsf.sty
\usepackage{graphicx}
\usepackage{epsfig}
\title[Spatial resolution effects on cluster studies]
{The effect of spatial resolution on optical and near-IR studies of stellar clusters: Implications for the origin of the red excess}

\author[N. Bastian et al.]{N. Bastian$^1$, A. Adamo$^{2,3}$, M. Schirmer$^4$, K. Hollyhead$^1$, Y. Beletsky$^{5}$, G. Carraro$^6$, \newauthor B. Davies$^1$,  M. Gieles$^{7}$, E. Silva-Villa$^{8,9}$\\
$^{1}$ Astrophysics Research Institute, Liverpool John Moores University, 146 Brownlow Hill, Liverpool L3 5RF, UK\\
$^{2}$ Department of Astronomy, Stockholm University, Oscar Klein Centre, AlbaNova, Stockholm SE-106 91, Sweden \\
$^{3}$ Max Planck Institut f\"{u}r Astronomie, K\"{o}nigstuhl 17, D-69117 Heidelberg, Germany\\
$^{4}$ Gemini Observatory, Casilla 603, La Serena, Chile \\
$^{5}$ Las Campanas Observatory, Carnegie Institution of Washington, Colina el Pino, Casilla 601, La Serena, Chile\\
$^{6}$ ESO, Alonso de Cordova 3107, Casilla 19001, Santiago de Chile, Chile \\
$^{7}$ Department of Physics, University of Surrey, Guildford GU2 7XH, UK\\
$^{8}$ (CRAQ) Universit\'e Laval, 1045 Avenue de la M\'edecine, G1V 0A6 Qu\'ebec, Canada\\
$^{9}$ FACom-Instituto de F'sica-FCEN, Universidad de Antioquia, Calle 70 No. 52-21, Medell\'in, Colombia
}
\date{Accepted. Received; in original form}
\pagerange{\pageref{firstpage}--\pageref{lastpage}}
\pubyear{2014}
\begin{document}
\maketitle
\label{firstpage}
\begin{abstract}
 Recent ground based near-IR studies of stellar clusters in nearby galaxies have suggested that young clusters remain embedded for $7-10$~Myr in their progenitor molecular cloud, in conflict with optical based studies which find that clusters are exposed after $1-3$~Myr. Here, we investigate the role that spatial resolution plays in this apparent conflict.  We use a recent catalogue of young ($<10$~Myr) massive ($>5000$~\msun) clusters in the nearby spiral galaxy, M83, along with Hubble Space Telescope (HST) imaging in the optical and near-IR, and ground based near-IR imaging, to see how the colours (and hence estimated properties such as age and extinction) are affected by the aperture size employed, in order to simulate studies of differing resolution.  We find that the near-IR is heavily affected by the resolution, and when aperture sizes $>40$~pc are used, all young/blue clusters move red-ward in colour space, which results in their appearance as heavily extincted clusters.  However, this is due to contamination from nearby sources and nebular emission, and is not an extinction effect.  Optical colours are much less affected by resolution.  Due to the larger affect of contamination in the near-IR, we find that, in some cases, clusters will appear to show near-IR excess when large ($>20$~pc) apertures are used.  Our results explain why few young ($<6$~Myr), low extinction ($\av < 1$~mag) clusters have been found in recent ground based near-IR studies of cluster populations, while many such clusters have been found in higher resolution HST based studies.  Additionally, resolution effects appear to (at least partially) explain the origin of the near-IR excess that has been found in a number of extragalactic YMCs.

\end{abstract}

\section{Introduction}
\label{sec:intro}

Due to the lower-extinction, as well as the near-IR focus of the next generation of ground and space based telescopes, the near-IR holds tremendous potential for stellar population studies.  While simple stellar population models predict that the colours of evolving stellar clusters do not vary strongly after $8-10$~Myr, the near-IR offers the opportunity to pick out highly embedded clusters, both through their stellar continuum as well as their nebular line emission (e.g., Br$\gamma$).  For example, mid-IR surveys such as the Spitzer Glimpse Survey (Churchwell et al.~2009) allowed the discovery of two (and subsequently many more) highly extincted massive young clusters located near the end of the Galactic bar (Figer et al.~2006; Davies et al.~2007).

Grosb{\o}l \& Dottori (2012 - hereafter GD12) have used near-IR imaging of a sample of relatively nearby ($10-25$~Mpc) spiral galaxies in order to study their cluster populations.  Their sample contained a number of objects\footnote{Due to the resolution of their images, the authors generally did not study individual clusters, but rather larger associations and cluster complexes, although we refer to them as ``clusters" here for simplicity.} that had red colours as well as Br$\gamma$-emission.  This led the authors to conclude that these clusters were young (ages $<$7~Myr) and  highly extincted ($\av > 3$~mag), meaning that they would likely be undetectable in the optical.  Additionally, few clusters with blue near-IR colours and Br$\gamma$-emission were found.  This result is in apparent contradiction with a number of optically based photometric and spectroscopic surveys that find a large number of clusters with ages less than 7~Myr (e.g., Bastian et al.~2009, 2012; Whitmore et al.~2010; Adamo et al.~2011a).  Additionally, studies of young massive clusters in the Milky Way and nearby galaxies suggest that clusters transition from being embedded in their natal molecular cloud to be ``exposed" in $1-2$~Myr (see Longmore et al.~2014 for a recent review).   A similar conclusion was reached by Whitmore \& Zhang~(2002), who compared catalogues of radio continuum sources (deeply embedded, very young clusters) and optically detected clusters, and found that the majority ($\sim85$~\%) of the optically detected clusters had radio counterparts, suggesting that the highly embedded phase lasts for only a few Myr.

Using HST J and H-band imaging (F110W and F160W), Gazak et al. (2013) found a number of optically detected massive clusters with ages less than $6$~Myr in the nearby spiral galaxy, M83.  These clusters have very blue near-IR colours, reflecting the lack of red-supergiants within them (c.f. Westmoquette et al.~2014).  Why are such clusters missing in other near-IR based cluster studies?

In addition to the above paradox, a number of recent studies have found an excess of flux in the near-IR, relative to that expected from simple stellar population models, in studies of extragalactic young clusters (e.g., Reines et al.~2008, 2010; Adamo et al.~2010a,b, 2011a,b).  Part of the excess comes from the nebular emission\footnote{Throughout this work we use ``nebular emission" to refer to both line and continuum nebular emission.} from the ionised gas around the clusters (Reines et al.~2010, Adamo et al.~2010a), which begins to significantly contribute to the integrated flux of the cluster from the I-band red-ward.  However, even when taking the nebular emission into account, a number of clusters show near-IR excess that increases with increasing wavelength (e..g, Adamo et al.~2010a).  The origin of this excess is still under debate, with potential contributions from nearby young stellar objects (YSOs), heavily extincted nearby massive clusters, or contamination from nearby red supergiants.

A potential caveat to most near-IR based studies that have been carried out so far, is the lower spatial resolution achievable from the ground relative to optical, HST based surveys.  For example, the study of GD12 of cluster complexes in a sample of grand design spiral galaxies, use an aperture radius of 0.5", corresponding to a physical radii of $23$ to $63$~pc at the distance of their targets ($9.5-26$~Mpc).  The authors acknowledge that their apertures do not cover just individual clusters, but rather contain cluster complexes.  In comparison, HST based optical studies typically have aperture sizes of 3-10~pc (e.g., Whitmore et al.~2010; Silva-Villa \& Larsen~2011).   Randriamanakoto et al.~(2010) have tested the role of resolution in determining the luminosity function index of cluster/complexes, and found that decreasing the resolution led to mildly flatter luminosity functions ($\sim0.1$ in the index), however the authors did not investigate how the lower resolution affected the derived cluster/complex properties.

The questions that we address in the current work are: what is the effect of resolution on optical and near-IR studies of extragalactic clusters?  Does contamination of surrounding objects affect optical and near-IR studies equally?  What is the cause of the apparent contradiction between near-IR and optically based studies, the former finding a lack of exposed young clusters and the latter finding many? 

In order to address these questions we make use of the ongoing study of the cluster population in the nearby ($4.5$~Mpc) face-on spiral galaxy, M83.  The proximity of this galaxy offers an excellent opportunity to compare results between different size apertures, to mimic the effect of distance/resolution in determining the cluster properties.  We use the optically based HST/WFC3 survey of 7 pointings across the face of M83, covering a large fraction of the optical extent of the galaxy, and the cluster catalogue presented in Silva-Villa et al. (2014), to search for (partially) embedded clusters.  The clusters were selected from V-band images, and had their ages, extinctions and masses estimated based on comparison between their U, B, V, H$\alpha$, and I-band magnitudes with simple stellar population models.  We refer to the above article for more information on the cluster sample.  We compliment this study with ground based {\em VLT-HAWK-I} J, H, K-band images, that allow us to perform photometry with apertures of $11$ to $87$~pc, in order to see how the magnitudes and colours change as a function of aperture size and wavelength.

This paper is organised as follows.  In \S~\ref{sec:obs} we present the data and techniques used and in \S~\ref{sec:results} we show our main results.  The implications of our findings are discussed in \S~\ref{sec:discussion}.

\section{Observations and Techniques}
\label{sec:obs}

\subsection{HST WFC3 Imaging}
\label{sec:hst}

For the present work we use archival HST/WFC3 images from programme IDs 11360 (PI O'Connell) and 12513 (PI. Blair).  The dataset consists of imaging with the F336W, F438W, F547M, F657N, F814W, F110W, and F160W filters.  We will refer to these filters as U, B, V, H$\alpha$, I, J, and H, respectively, although no transformations were applied.  For one of the fields (Field~1) the F555W filter was used instead of the F547M filter.  The F110W filter images are only available for Fields 1 and 2.  Further details on the data used are given in a future work, Silva-Villa et al.~(2014).

\subsection{Cluster catalogue}
\label{sec:clusters}

In order to test the effects of resolution on cluster studies, we use the catalogue of clusters and associations from Silva-Villa et al.~(2014).  In that work, clusters were identified on seven V-band WFC3 HST pointings that cover a large fraction of the optical extent of the galaxy.  Based on visual inspection, sources were labelled as either class 1 (likely clusters, centrally concentrated, resolved objects) or class 2 (likely associations, multiple centres, highly elongated).  Ages, masses and extinctions were estimated through a comparison of each sources U, B, V, H$\alpha$ and I-band magnitudes with simple stellar population models (e.g., Adamo et al. 2010a).  In the present work, we use the Silva-Villa et al. catalogue and select class 1 and 2 sources with ages $\lesssim10$~Myr and masses $>5000$\msun.  The mass cut was applied in order to limit the effects of stochasticity of the stellar IMF in the broad band properties of the sources (e.g., Barbaro 
\& Bertelli~1977; Fouesneau \& Lan{\c c}on~2010; Silva-Villa  \& Larsen~2011; de Meulenaer et al.~2013).

\subsection{VLT HAWK-I Imaging}
\label{sec:vlt}

{\em VLT/HAWK-I} covers a field of view of $8^\prime\times8^\prime$ with a pixel 
scale of 0.104", using a mosaic of 4 HAWAII-2 detector arrays.
In order to cover M83, a $5^\prime$ wide dither pattern was chosen. Hence 
the central $4^\prime-5^\prime$ of M83 received 4 times more integration 
time than its outskirts. Data were taken over 11 nights between 2009-01-02 
and 2009-03-25 (ESO program ID 382.D-0181; PI Gieles). Blank sky fields were 
interspersed for background subtraction.

The source density of common astrometric reference catalogs in this area is 
insufficient for full distortion correction and image registration. We thus 
worked with a secondary reference catalog from archival $R$-band data taken 
in good seeing with the Wide Field Imager at the 2.2m MPG/ESO telescope 
(program ID 69.C-0426, PI: Alves). A common astrometric solution was derived 
for all {\em HAWK-I} exposures, with an internal uncertainty of $\sim$1/5th pixel.

Background subtraction with separate blank fields removed most of the sky 
signal. However, small non-zero offsets remain in the individual exposures. 
These were corrected for by measuring a constant sky value in a blank area 
of one of the 4 chips of the M83 pointings. All data reduction steps were 
done with THELI (see Schirmer 2013 for details, and Erben et al. 2005).

The images were calibrated off of 2MASS (Skrutskie et al.~2006) photometry, using $\sim10$ bright and isolated sources per filter, per pointing.  Aperture photometry for the $\sim10$ sources (per image/filter) were carried out with apertures of 5, 10, 20 and 40 pixels.  These sources consisted of a mixture of stars and isolated clusters (i.e., resolved in HST images).  Due to the seeing, all sources appeared point-like and no difference were observed in the zero point derived from the clusters or stars, even for the smallest apertures (5 and 10 pixels).  The zero points for the 20 and 40 pixel apertures were derived simply by comparing the observed magnitudes with the 2MASS magnitudes.

For the zero points for the smaller apertures (5 and 10 pixels) we also needed to correct for differences in the seeing for the different filters/images (see Table~\ref{hawki_data}).  For this, we used the additional information on the extent of the spatial profile (i.e., the seeing during that observation), by looking at the magnitude difference between a 2 and 5 pixel aperture.  As above, we adopted the magnitude of the sources from 2MASS as the standard. 

The final magnitude for the 5~pixel aperture was then calculated as follows:
$${\rm Mag}_{\rm 5, final} = {\rm Mag}_{\rm 5, observed}  + {\rm Mag}_{\rm 5, correction},$$
where 
$$ {\rm Mag}_{\rm 5, correction} = {\rm c}_1 * ({\rm Mag_{5,observed} - Mag_{2,observed}}) + {\rm c_2}, $$
and ${\rm c_1}$ and ${\rm c_2}$ are constants derived from all of the 2MASS sources used for calibration per filter ($\sim40$ isolated sources).  The same approach was used for the 10~pixel apertures.

Using this method, the standard deviation of the final adopted magnitude, relative to the 2MASS magnitude, was $\sim0.06$~mag for each filter and aperture size. 

\begin{table}
\caption{Characteristics of the VLT/HAWK-I data set. Since the data were 
taken over 11 different nights, and a wide dither pattern was used, the 
image seeing is not constant across the field.}
\label{hawki_data}
\begin{tabular}{lcc}
\noalign{\smallskip}
\hline
\hline
\noalign{\smallskip}
Band & $t_{\rm exp}$ [s] & Seeing \\
\hline
\noalign{\smallskip}Þ
$J$ & 20$\times$60s  & 0.36$-$0.67"\\
$H$ & 60$\times$60s  & 0.33$-$0.68"\\
$K$ & 48$\times$60s  & 0.37$-$0.72"\\
$BrG$ & 16$\times$300s & 0.37$-$0.59"\\
\hline
\end{tabular}
\end{table}

\section{Results}
\label{sec:results}

\subsection{Near-IR colours}
\begin{figure}
\includegraphics[width=8.5cm]{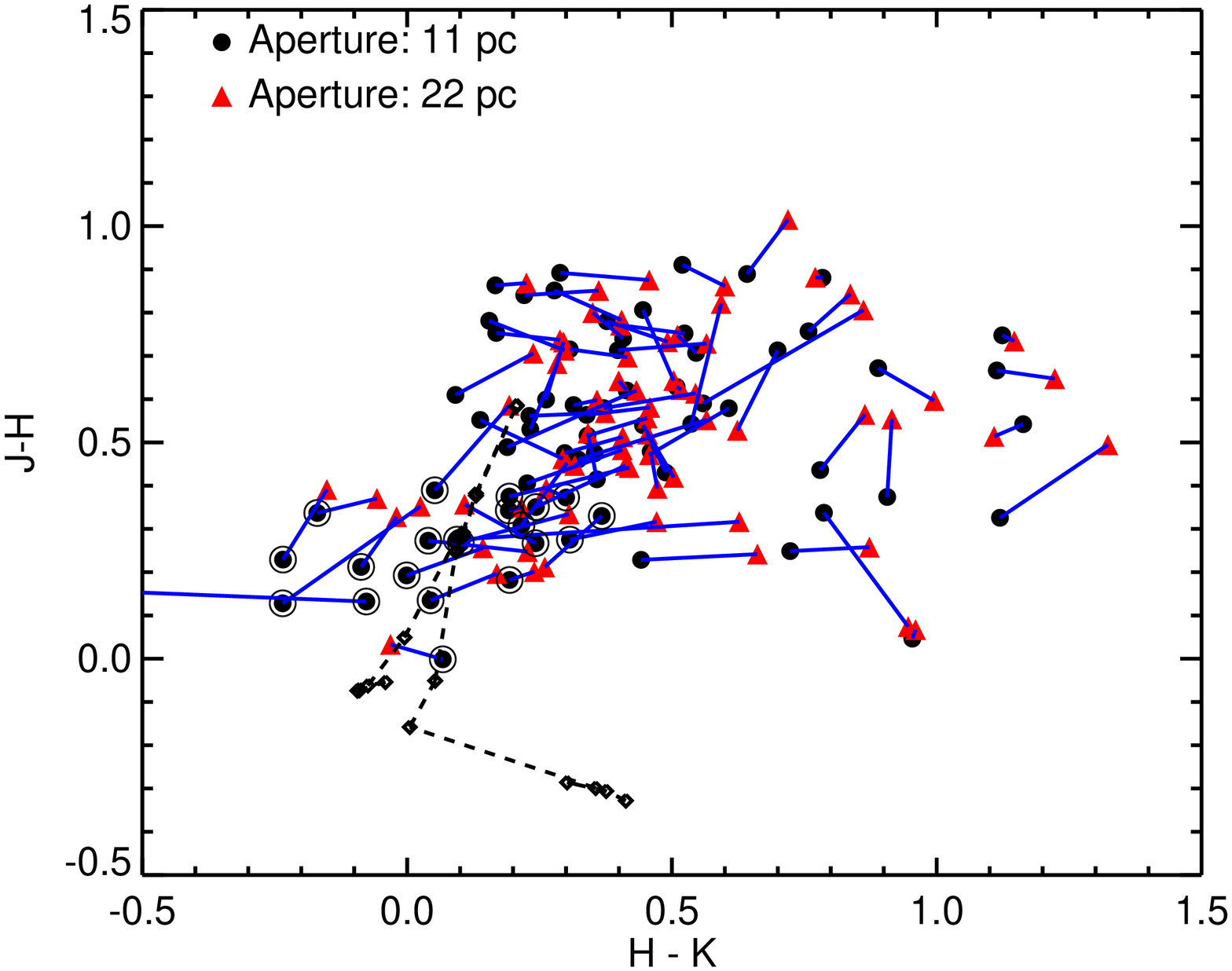}
\includegraphics[width=8.5cm]{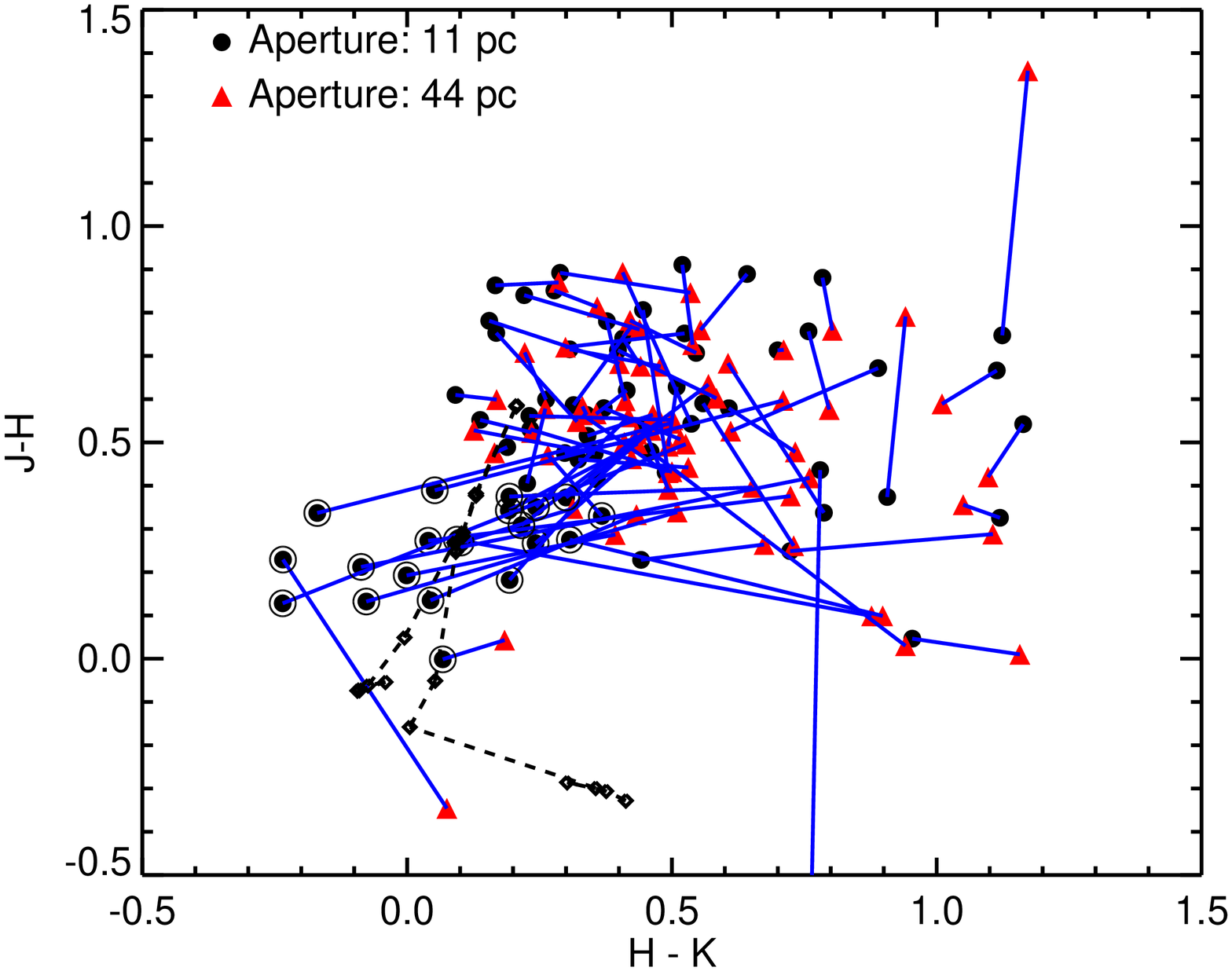}
\includegraphics[width=8.5cm]{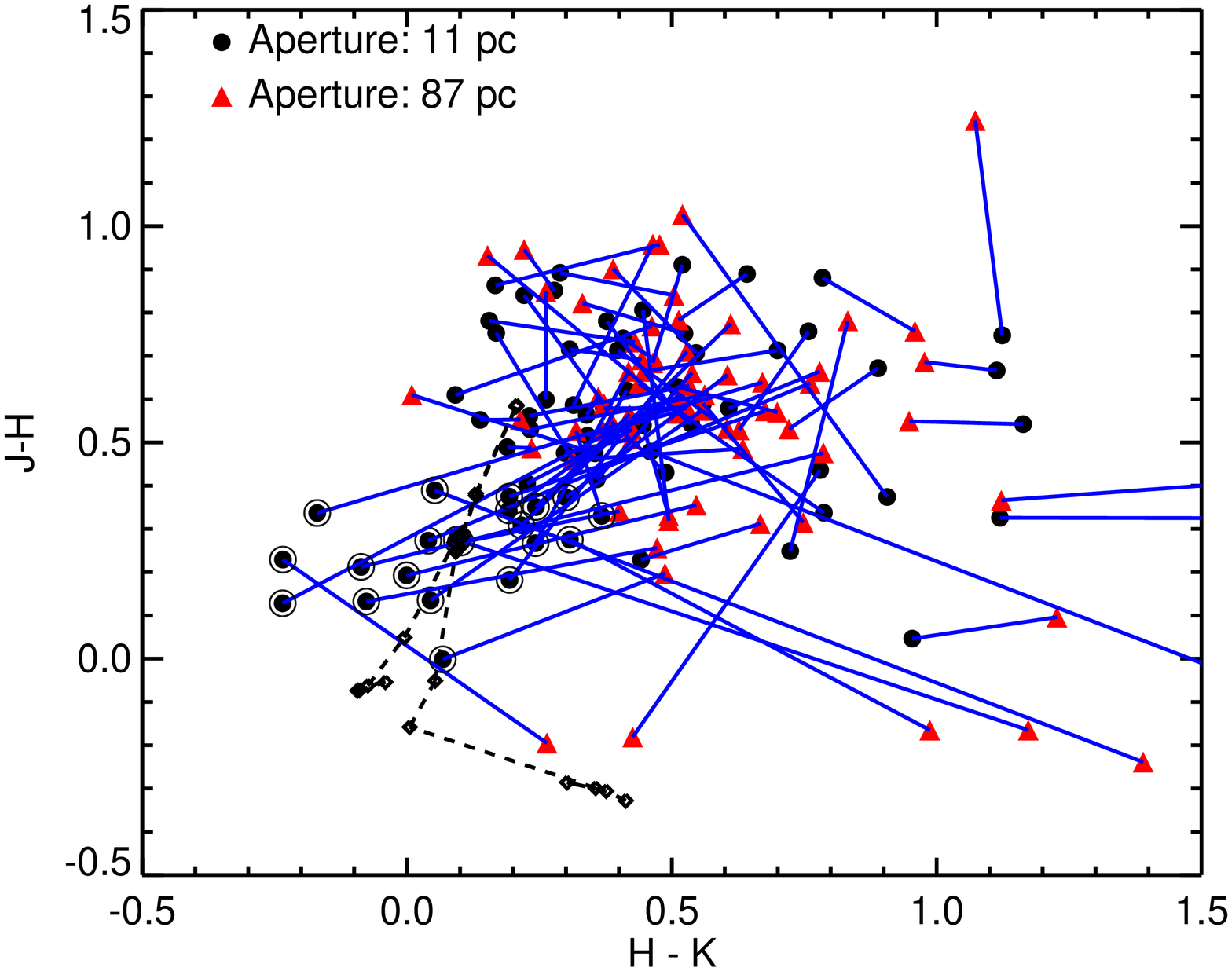}
\caption{The effect of aperture size on the resulting near-IR colours of young ($\lesssim10$~Myr) clusters.  The black dots  show the colour of clusters when a $\sim11$~pc aperture is used. In the upper panel we highlight blue clusters with open circles.  Red dots represent the same clusters but now measured with larger apertures, connected by solid (blue lines).   Note that the blue clusters become redder and join the main (i.e. the ``red cloud") locus of points as larger apertures are used.}
\label{fig:cc_nir}
\end{figure}

\begin{figure}
\includegraphics[width=8.5cm]{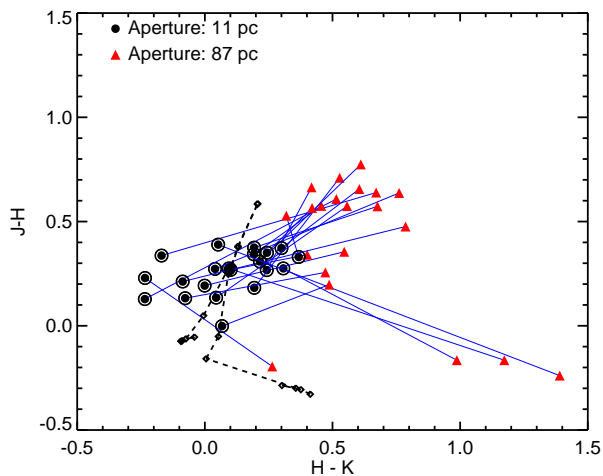}
\caption{The same as the bottom panel of Fig.~\ref{fig:cc_nir2} except here, for clarity, only clusters with blue colours (when a $\sim11$~pc aperture is used), are shown.}
\label{fig:cc_nir2}
\end{figure}

Figure~\ref{fig:cc_nir} shows the $H-K$ vs. $J-H$ colours of our cluster sample for different aperture sizes.  In each panel, the $11$~pc aperture is shown as filled black circles, and a larger aperture is shown with filled red triangles, and the two points are connected with solid (blue) lines in order to see how each cluster is affected by the aperture size.  Additionally, in each panel we highlight the clusters with blue near-IR colours with open circles.  These are the young, relatively low extinction clusters (some examples are shown in Fig.~\ref{fig:cc_optical}).  Note how these blue clusters move to redder colours as larger apertures are used.  The dashed lines (with diamonds) show the Yggdrasil simple stellar population models (SSP) (Zackrisson et al. 2011), from 1 to 10~Myr (from low to high $J-H$ colour) for solar metallicity.  Two models are shown, one including nebular (continuum and line) emission (the redder model in $H-K$) and one of pure stellar continuum.  Additionally, in Fig.~\ref{fig:cc_nir2} we show the same as in the bottom panel of Fig.~\ref{fig:cc_nir}, however only including the clusters with blue near-IR colours (circled).  The red-ward trend is clearer in this representation.

For the blue clusters in the $11$~pc aperture (those circled), note how they move to the red as larger apertures are used.  For aperture sizes of $44$~pc, nearly all of them have moved to the ``red cloud", centred at $H-K=0.5$ and $J-H=0.6$.  Hence, studies that use apertures of this size would not be expected to find blue clusters.  

\begin{figure*}
\begin{center}
\includegraphics[width=3cm]{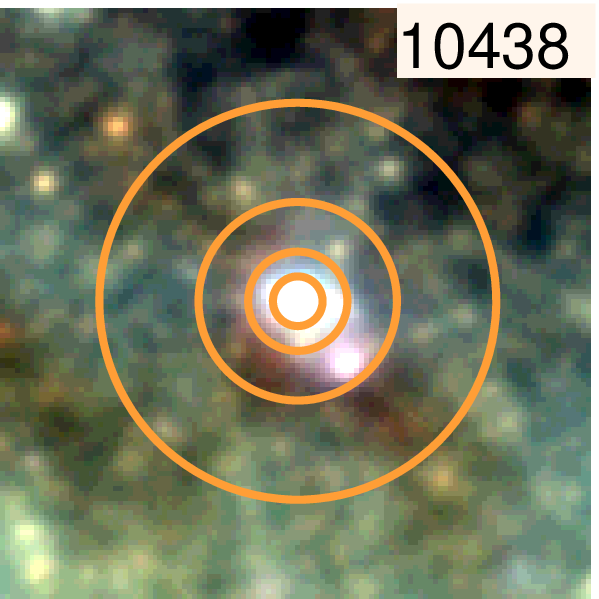}
\includegraphics[width=3cm]{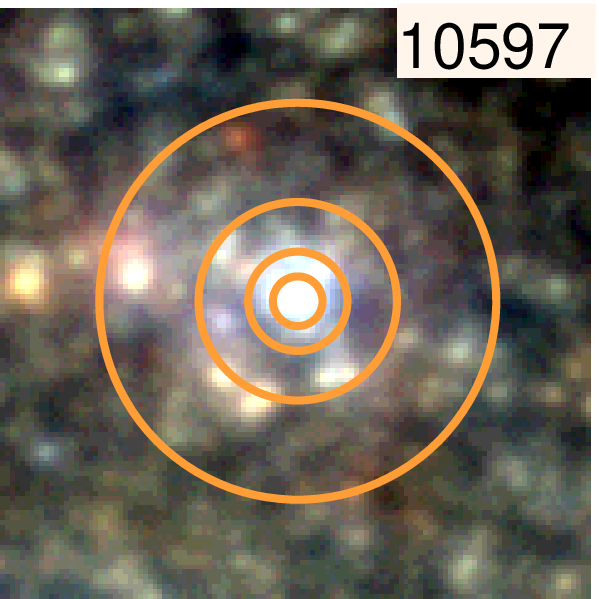}
\includegraphics[width=3cm]{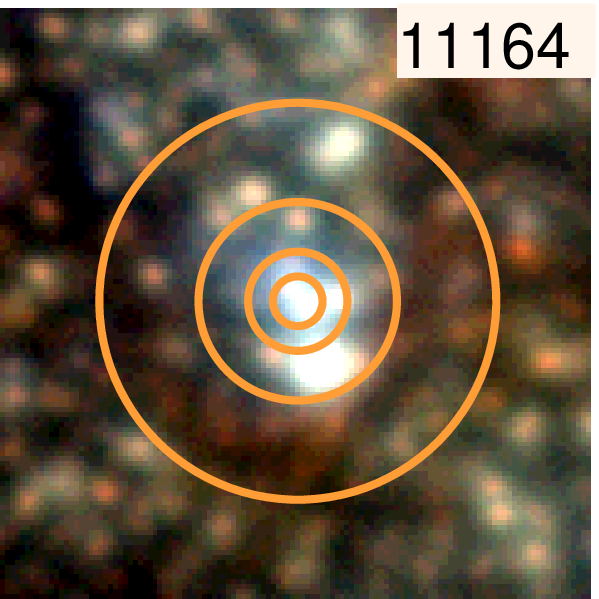}
\includegraphics[width=3cm]{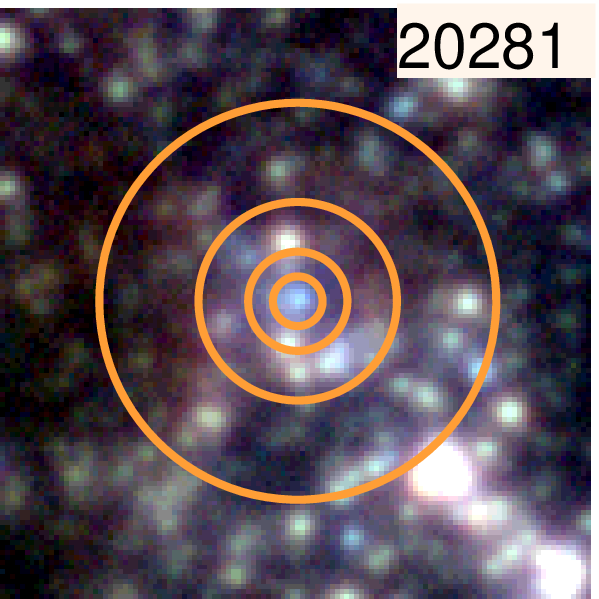}
\includegraphics[width=3cm]{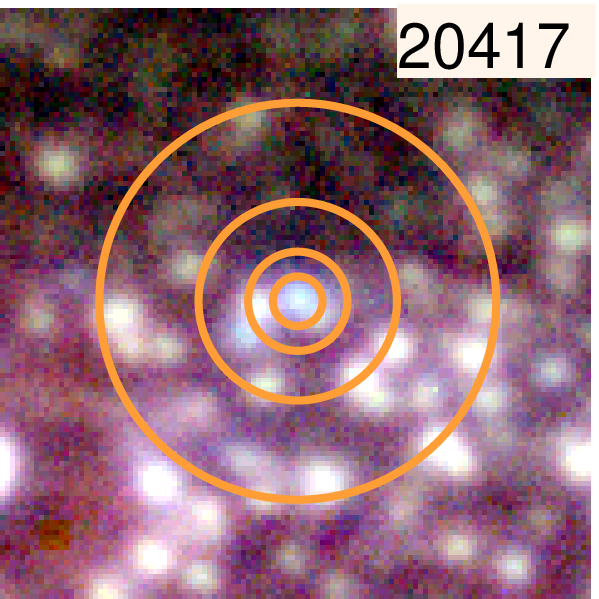}
\includegraphics[width=3cm]{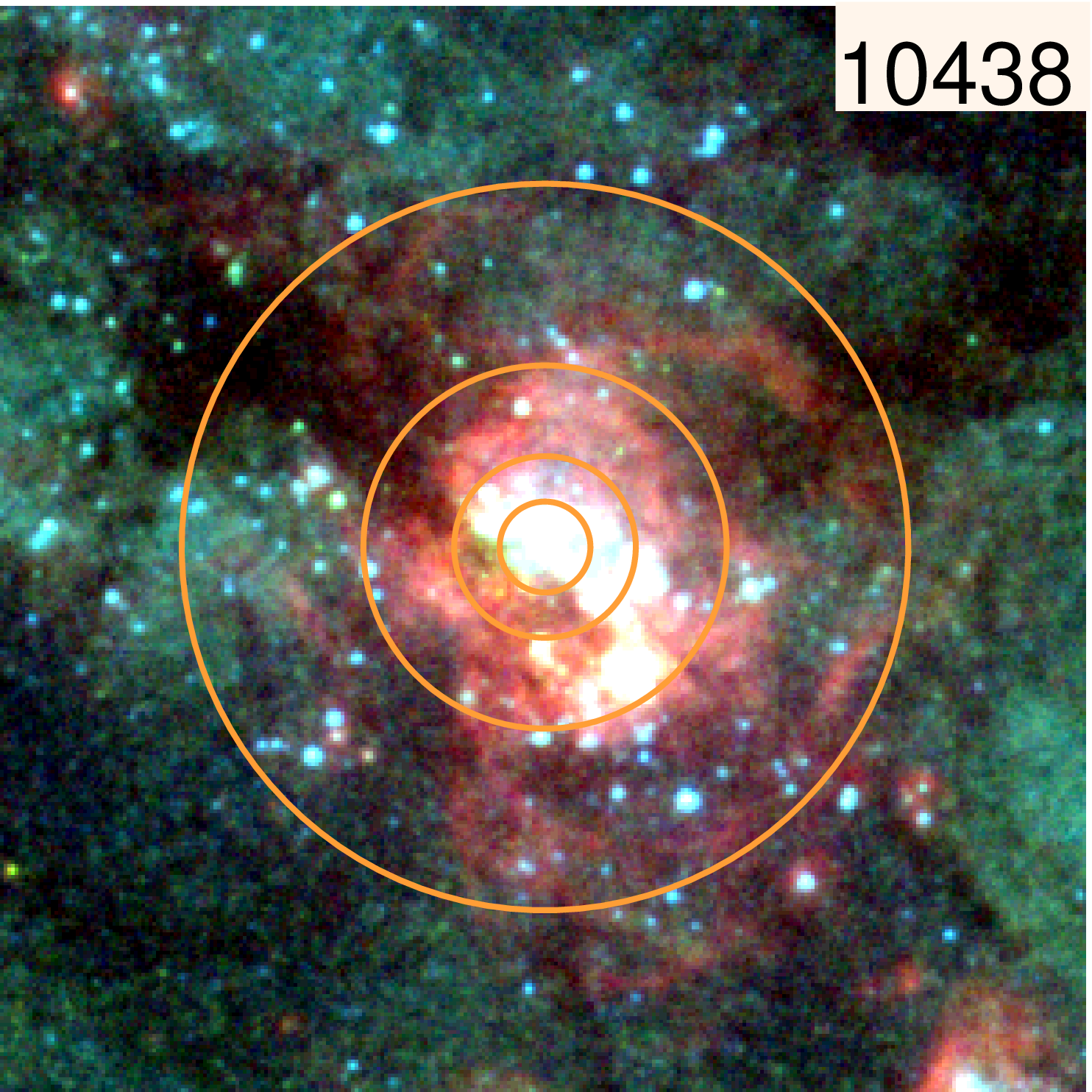}
\includegraphics[width=3cm]{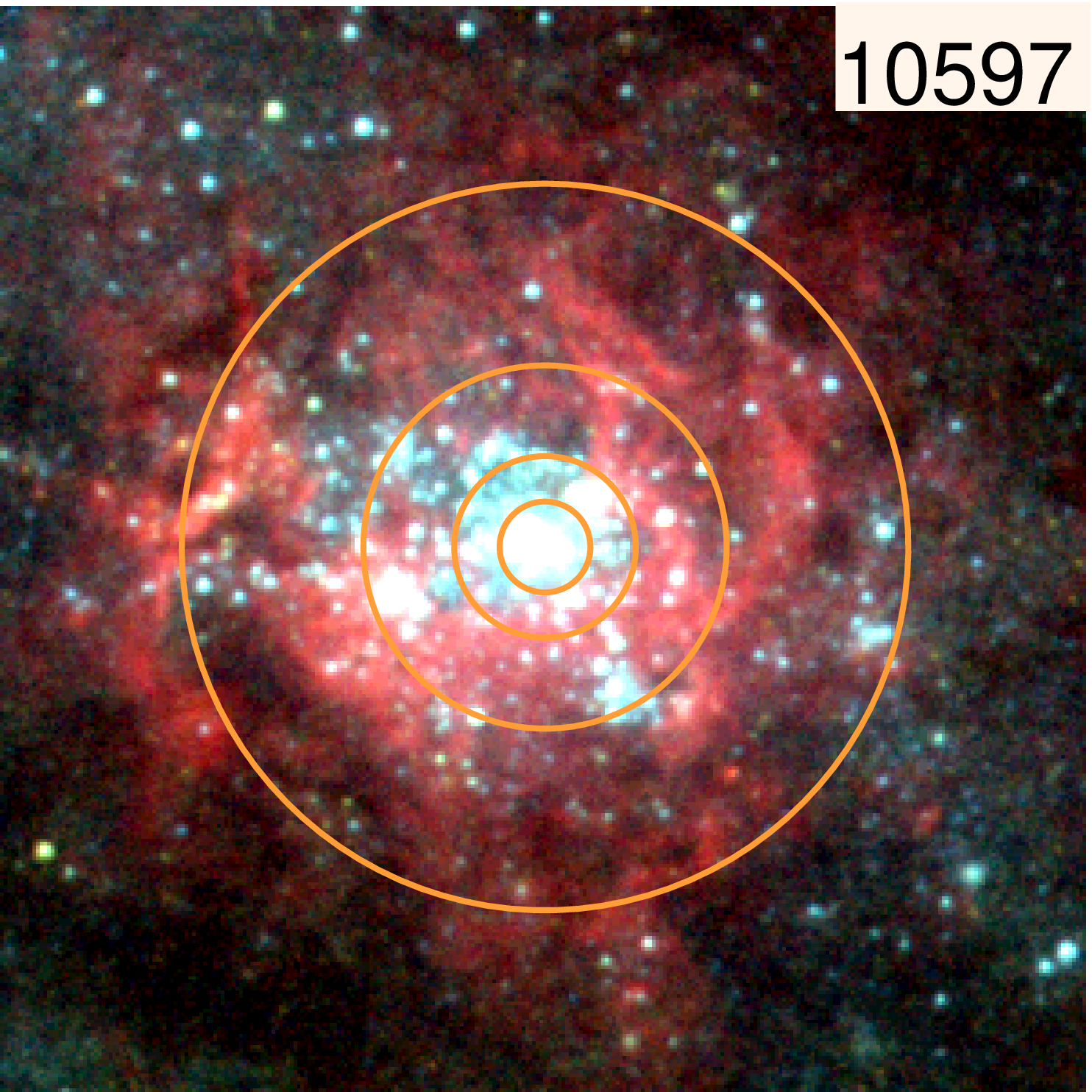}
\includegraphics[width=3cm]{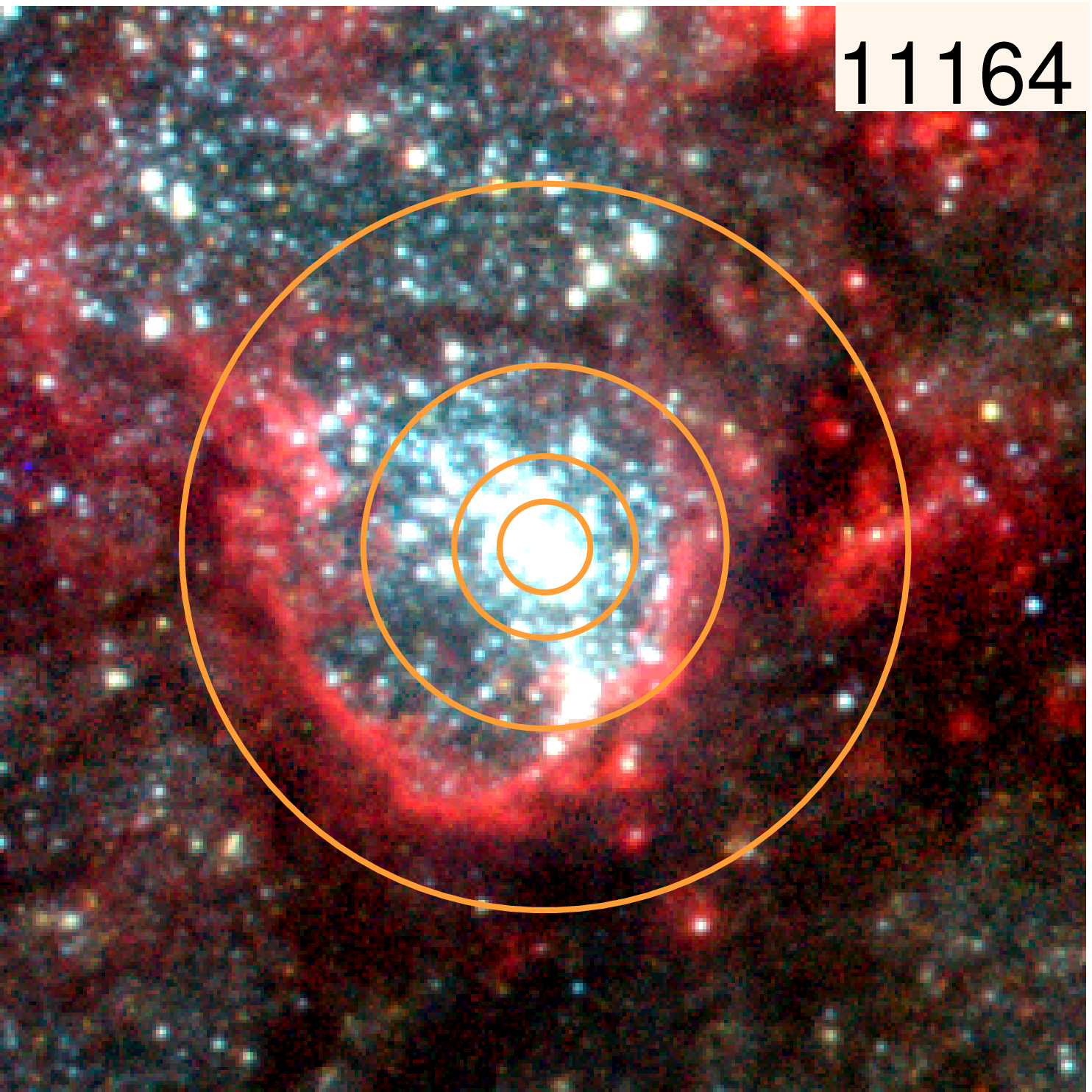}
\includegraphics[width=3cm]{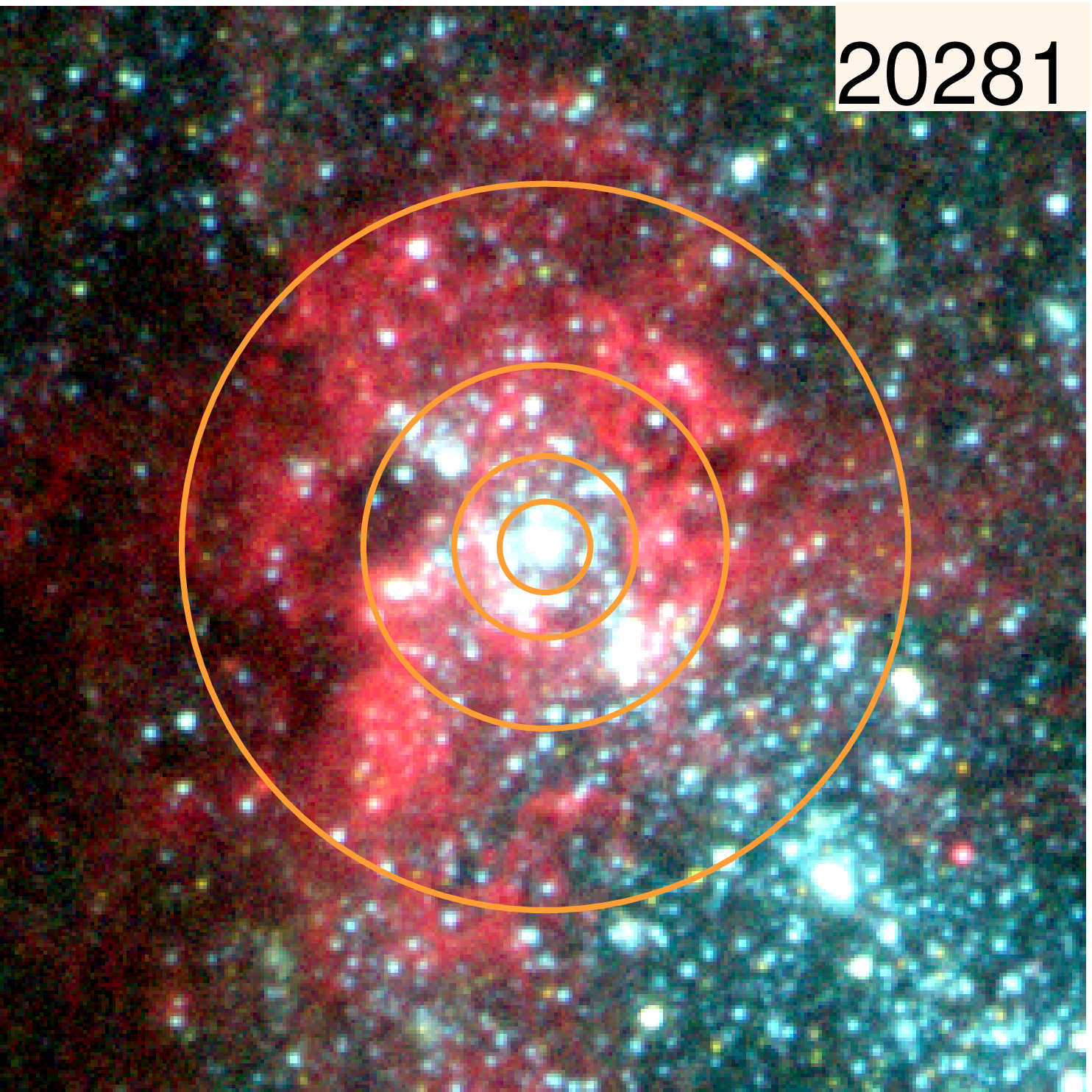}
\includegraphics[width=3cm]{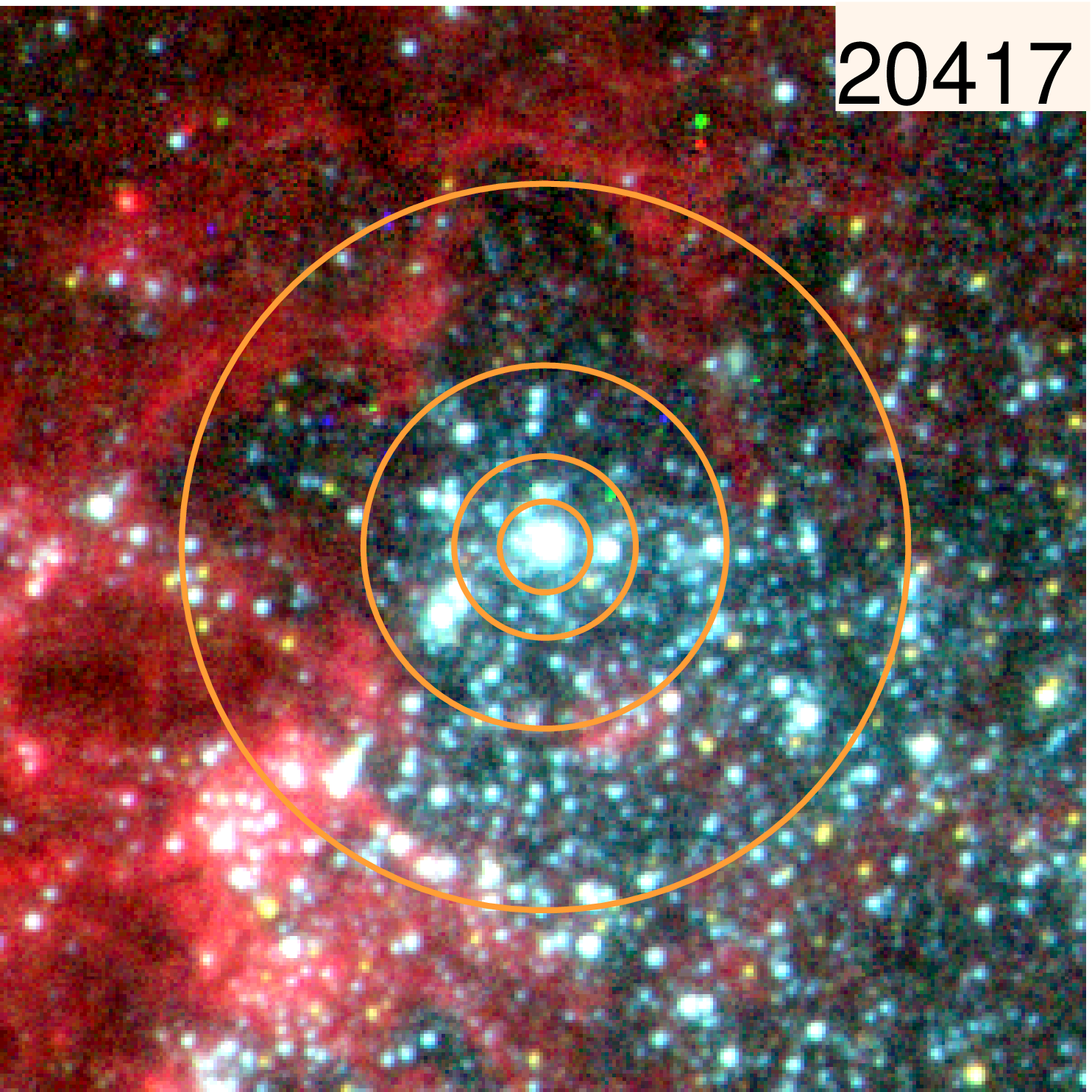}
\includegraphics[width=17cm]{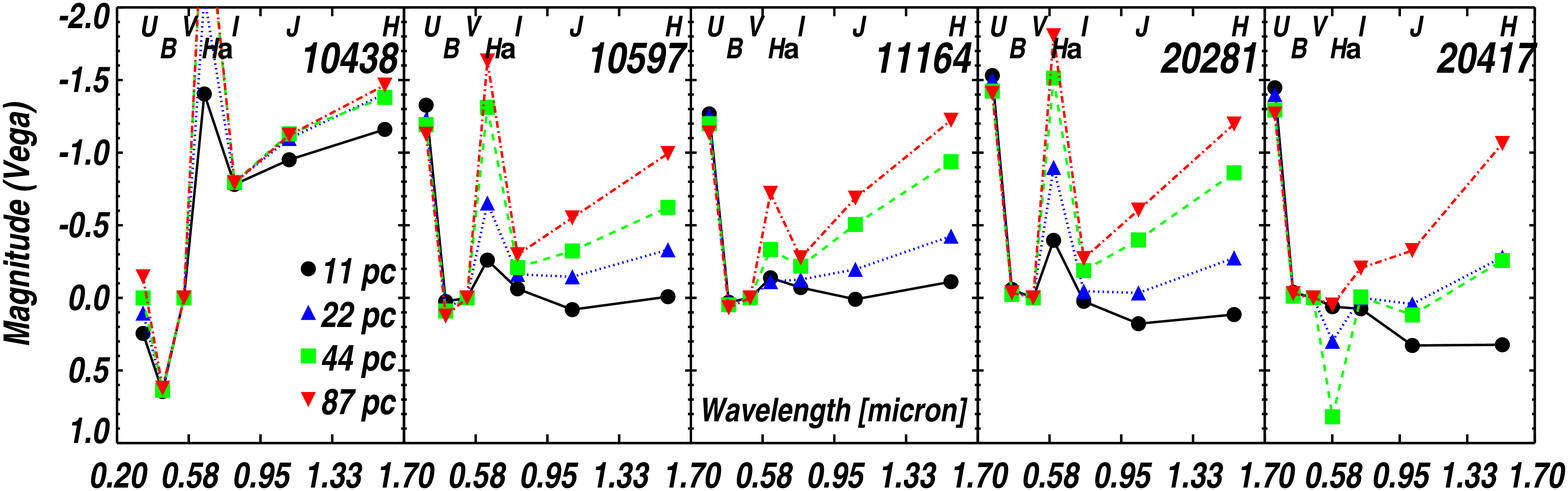}
\caption{False colour composite images of a sample of clusters used in the current work. {\bf Upper panels:} HAWK-I J, H, K-band composites.  {\bf Middle panels:} HST WFC3 B, V, H$\alpha$-band composites. The apertures used (11, 22, 44, 87~pc) are shown.  In most  of the near-IR images, a source with a brightness similar to or larger than the central source is contained within the 44~pc aperture.  However, in the optical images, the central source often remains the dominant source even within the 87~pc aperture.  {\bf Lower panels:}  The HST based (including J and H-band) spectral energy distribution for the five sources above for different aperture sizes, each has been normalised to the flux in the V-band.  Note that for larger aperture sizes, the flux in H$\alpha$ increases strongly, due to picking up the emission outside the hole that has formed around the central massive cluster.  Also, note that all clusters display significant near-IR (beginning in the I-band) excess as larger apertures are used, while the optical/UV magnitudes remain largely unchanged.}
\label{fig:images}
\end{center}

\end{figure*}

False composite colour images of a sample of clusters and their surroundings are shown in Fig.~\ref{fig:images}.  These clusters were chosen to have blue near-IR colours (when small apertures were used), and also to display a range in H$\alpha$ morphology.  Such morphology will be discussed in more detail in Hollyhead et al.~(in prep.).  In the upper panel we show J, H, K-band HAWK-I colour composites, whereas in the middle panels HST/WFC3 B, V, and H$\alpha$ composites are shown.  For the near-IR composites, we note that most clusters have a source within 44~pc with a brightness that is similar to, or higher than, the central cluster.

We note that a $\sim23$~pc radius is the smallest aperture used in the GD12 study.  This explains why young clusters in their survey (those with Br$\gamma$ emission) have red colours.  It is not due to extinction, but rather contamination from nearby sources that are bright in the near-IR as well as the nebular emission from the ionised shell (that is often seen around young clusters with sizes between $10-50$~pc - Hollyhead et al.~in prep.).

Additionally, we explored the effect of adopting different annuli surrounding the clusters in order to subtract the background flux.  For older clusters ($>10$~Myr) the measured colours and magnitudes were not strongly affected by the size of the annuli.  For younger ($<10$~Myr) clusters we found the following results.  As expected, the measured magnitudes and colours displayed the least scatter when small annuli ($<20$~pc) were used.  For annuli larger than $100~pc$ we also found that the magnitudes and colours displayed small scatter (i.e. differences between using annuli of 100 and 150~pc).  For intermediate annuli ($20-100$~pc) the scatter was the largest.  The reason for this is that young clusters tend not be born in isolation, but rather often have surrounding clusters and/or young field populations around them (e.g., Bastian et al.~2005).  Hence, whether or not a bright nearby source (or a number of them) enter the background annuli can have a significant effect on the measured colours and magnitudes.

While the main goal of the present work is to explore the effect that resolution plays in the study of young clusters, we also investigated the effect on older clusters.  To do this, we carried out similar tests on a sample of clusters with ages between 20 and 200~Myr, taken from the Bastian et al.~(2012) sample.  We found that for these clusters the near-IR (and optical) colours were not dependent on the aperture size used.  The reason for this is that these clusters are relatively isolated, and are no longer associated with their highly structured natal star-forming region.  These star-forming regions dissolve into the field on $\sim10$~Myr timescales (e.g., Gieles \& Portegies Zwart~2011), so finding an older cluster ($>20$~Myr) near a younger ($<10$~Myr) cluster is much rarer than finding a young cluster ($<6$~Myr) near a slightly older cluster ($6-10$~Myr).  Hence, we conclude that the young ($<10$~Myr) regions will be much more affected by the spatial resolution than older clusters ($>20$~Myr).

\subsection{Optical colours}

\begin{figure}
\includegraphics[width=8.5cm]{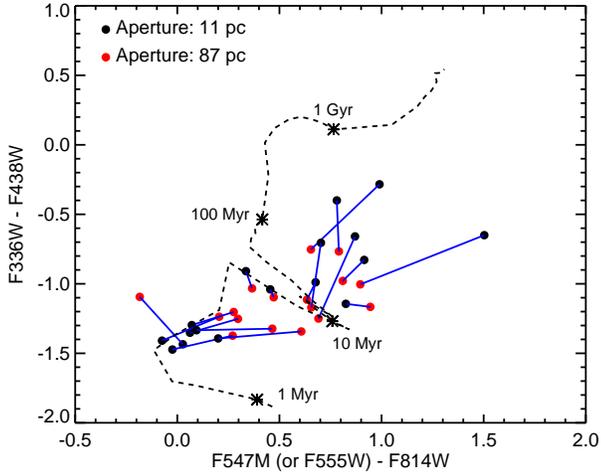}
\caption{Similar to Fig.~\ref{fig:cc_nir} but now for optical colours.  We only show the most extreme case (apertures of 11 and 87~pc) as even in this extreme case, it is clear that the colour changes would not be enough to significantly shift the best fitting age and extinction.}
\label{fig:cc_optical}
\end{figure}

We carried out a similar experiment in the optical and show the results in Fig.~\ref{fig:cc_optical}.  Here we only show the results for the $11$ and $87$ radius apertures, as even at these extremes, it is clear that the effect of resolution is much less in the optical than in the near-IR.  Additionally, we are only showing those clusters which appear blue in the $11$~pc aperture near-IR colour-colour plot (i.e., those circled in the top panel Fig.~\ref{fig:cc_nir}). The youngest, non-extincted clusters ($<6$~Myr, $V-I=0.1$, $U-B=-1.3$.) are moved to redder colours, but only by a relatively small amount, meaning that they would still be found as young, simply with a slightly higher extinction.  Clusters with relatively high extinction (\av$>1-2$~mag) and young ages become bluer when larger apertures are used.  However, this is only a modest change, so their position in colour-colour space would still lead to the same conclusion, i.e. that they are young clusters with high extinction.  

In order to quantify this, we fit all the clusters shown in Fig.~\ref{fig:cc_optical} to SSP models using the method discussed in Adamo et al.~(2010a,b,2011a,b; see also \S~\ref{sec:sed_modelling}) to estimate the age and extinction of each cluster, using only the optical photometry measured in 11 and 87~pc apertures.  The median age difference between the two fits was 0.08 dex (and a standard deviation of 0.2~dex) and the median difference in \av\ of 0.2 mag (and a standard deviation of 0.25 mag). By using the larger apertures, the mass of the cluster was overestimated by 0.8~dex, highlighting the significants amount of flux coming from nearby sources when studying young ($<10$~Myr) clusters.  Hence, we conclude that the estimated age and extinction of the clusters, based on optical and UV photometry, are not strongly affected by aperture size.

The use of larger apertures does result in (at least) one clear change in the optical spectral energy distribution (SED), namely the strength of emission lines.  As seen in the middle panels of Fig.~\ref{fig:images}, the young clusters often clear out a hole in the surrounding ISM, due to the combination of stellar winds, SNe, and their ionising flux.  These combine to drive an ionised shell into the surrounding ISM.  When the apertures cover the ionised ring (the edge, in projection, of the 3D shell) the amount of flux in emission lines goes up drastically.

\subsection{The origin of the infrared excess}

As discussed above, the near-IR colours and magnitudes are found to be more strongly affected by neighbouring sources and nebular emission than optical colours.  This could explain (at least partially) the origin of the near-IR flux excess (relative to SSP models with or without the inclusion of nebular emission) that has been found in a number of studies of young extragalactic clusters (e.g., Reines et al.~2008, Adamo et al.~2011a).  As seen in the bottom panels of Fig.~\ref{fig:images}, all the clusters begin to display significant amounts of near-IR excess when apertures greater than 20~pc are used.  We note that the excess near-IR emission found by Reines et al. (2008) is largely due to nebular emission (Reines et al. 2010), whereas Adamo et al. (2010, 2011a,b) used models that included nebular emission, suggesting that the excess near-IR found in those studies has a different origin.

\begin{figure}
\includegraphics[width=8.5cm]{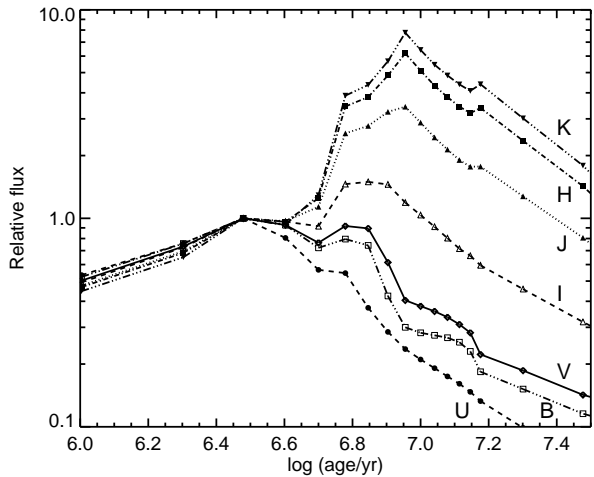}
\caption{Predictions from the {\em Yggdrasil} SSP models of the relative flux (normalised to the flux at an age of $3$~Myr) in different bands as a function of age.  Note that the flux in the near-IR (JHK) increases strongly after $\sim5$~Myr, whereas the optical (UBV) drops.  Hence, if an older cluster (or stellar population) is within a given aperture, its affect will be much stronger in the near-IR than the optical (i.e., for a similar mass the second cluster will dominate the near-IR while the first cluster will dominate the optical). }
\label{fig:flux_vs_age}
\end{figure}

In Fig.~\ref{fig:flux_vs_age} we investigate the cause of this by looking at the predicted flux in different bands as a function of age for a simple stellar population (we use the {\em Yggdrasil} SSP models for solar metallicity, and without nebular emission).  The models are normalised to the flux at an age of $3$~Myr.  After $\sim5-6$~Myr, when the RSGs begin to appear, we see a pronounced increase in the flux in the I, J, H, and K filters, while the U, B and V fluxes drop.  Hence, in the case of a young cluster, aged $\sim3$~Myr (the primary population), which has neighbouring source of similar mass with an older age $5-30$~Myr (the secondary population), the combined flux will be dominated by the primary population in the optical and the secondary population in the near-IR.

In addition to contamination from neighbouring stellar sources, another potential source of the infrared excess is the contribution from the nebular emission, including both the line and continuum (e.g., Reines et al.~2010).  As young clusters emerge from their natal cocoon, their collective stellar winds, SNe, and ionising flux drive a shell of ionised material away from the cluster (see examples shown in Fig.~\ref{fig:images} and also Whitmore et al.~2011).  While the cluster is still partially embedded, it should show the red-excess phenomenon (relative to SSP models that do not include nebular emission) even with small apertures.  As the shell expands, the nebular emission will drop in small apertures.  However, as the aperture increases the contribution of the nebular can increase rapidly as the ionised shell enters the aperture.  Hence, nebular emission also moves the clusters from blue near-IR colours to the ``red cloud" observed in Figure.~\ref{fig:cc_nir}.

A useful comparison can be made with the 30~Doradus region in the LMC.  In addition to the massive ($\sim6 \times 10^4$\msun) young ($2-3$~Myr) cluster, R136 (e.g., Portegies Zwart et al.~2010), that powers the majority of the surrounding H{\sc ii} region, a number of older clusters (and stellar populations in the field) are nearby.  One of these clusters is Hodge~301, a $\sim20-25$~Myr, $\sim5000$~\msun\ cluster located $\sim40$~pc away from R136 in projection, which contains a number of red supergiants (Grebel \& Chu~2000).  Based on the models shown in Fig.~\ref{fig:flux_vs_age}, Hodge~301 would be expected to contribute 50\% as much flux as R136 in the K-band, and around 1\% as much in the U-band. Additionally, the 30 Doradus region is known to host stellar populations (both between clusters and within the field) with significant age differences (tens of Myr - e.g., Walborn \& Blades~1997).  These populations would contribute significantly to the flux of R136 if apertures greater than $\sim20$~pc were used.  The above example highlights the role that other nearby clusters may have on the estimated properties of a massive clusters if a large aperture is used, however, a substantial field population (e.g., individual RSGs or AGBs) would have the same result.  Efremov \& Elmgeen~(1998) have shown that the age spread present in a region scales with the size of the region under study, so larger apertures will naturally sample larger age spreads within a region.

Most studies that have found the near-IR flux excess are based on distant galaxies, where large (physical) apertures are required due to resolution restrictions (e.g., Reines et al.~2008, Adamo et al.~2011b who used apertures of $\sim40$ and $\sim35$~pc, respectively).  Based on the results presented here, it appears that the basic properties of the clusters (which are estimated largely from optical colours) such as age and extinction are likely to be representative of the dominant cluster. However, the near-IR integrated flux may be contaminated by neighbouring sources and nebular emission. This excess flux accounts for some fraction of the observed near-IR excess, the exact amount will depend on the surroundings of the clusters.  In a future work, we will directly address the source of the NIR excess using high-spatial resolution VLT/SINFONI near-IR integral field spectroscopy of very young star clusters in M83 and Haro~11 (Adamo et al., in prep.).  However, of the five clusters shown in Fig.~\ref{fig:images}, three display strong H$\alpha$ emission and the red excess phenomenon when large apertures are used.  

\subsection{SED modelling}
\label{sec:sed_modelling}

We have carried out SED fitting for the five clusters shown in Fig.~\ref{fig:images} using the same method and SSP models ({\em Yggdrasil}, solar metallicity, Kroupa~2001 stellar IMF) as in Adamo et al. 2010a,b, 2011a,b), for each of the apertures used.  The models include nebular emission (both line and continuum emission), and we assume that 50\% of the ionising photons are absorbed and re-emitted within the aperture.  We only used the observed fluxes in the optical for the fits, and then compared the predicted J (F110W) and H (F160W) fluxes of the best fitting model to the HST observations.  

In general, we find that for the smallest (11~pc) apertures, the observed J and H-band fluxes are significantly fainter than the predictions.  This is due to the fact that the young clusters (as seen in Fig.~\ref{fig:images}) have cleared most of the natal gas cloud away (at least to radii larger than the aperture), meaning that significantly less gas is present than the model assumes, leading to lower emission (both in the continuum and line emission) than the model predicts.  As the apertures become larger (22~pc), the predicted and observed near-IR fluxes become comparable.  For the (at least partially) embedded cluster in Fig.~\ref{fig:images} (10438) the observed vs. predicted fluxes in the near-IR agree for the largest apertures (87~pc), suggesting that at this radius, roughly half of the ionising photons are absorbed.  For the other clusters, the largest apertures have significant amounts of excess emission above the model fit, showing that neighbouring sources are affecting the photometry.

The extreme sensitivity of the near-IR (and the optical emission lines) on the adopted 'covering factor' of the emission (i.e., the fraction of ionising photons absorbed) shows that care must be taken when including these filters in broad-band fits for stellar population properties.   The test performed on the M83 clusters suggests that, in some cases, when only the optical bands are fitted, our adopted models (which include stellar and nebular emission) are not able to reproduce the observed near-IR fluxes in larger apertures (i.e., a near-IR excess is present). 

Hence, there are two effects that contribute to the near-IR excess, nebular emission and nearby contaminating stellar populations.  Quantifying the amount that each type contributes is difficult.  Even models that take nebular emission into account may miss the actual flux, as they need to assume a covering fraction (i.e. the fraction of ionising photons that are absorbed within the aperture used).  Throughout this work, we have adopted the 50\% covering factor models to estimate the cluster parameters.  From Fig.~\ref{fig:images} we can see that some clusters have blow ionised bubbles around them, meaning that the covering fraction is lower than assumed in the centre (i.e. for small apertures) but may be appropriate for larger apertures.  Still other clusters appear to have ``blow outs", where a large fraction of the ionising photons may be escaping to large distances.  In principle, by modelling the SED with multiple emission lines included, one may be estimate the fraction of ionising photons that are escaping as a function of radius.  In practice, however, such modelling is limited by the uncertainties in the ionising photon output of massive star models, making the conversion between the observed emission line strength to the ionising photon escape fraction highly uncertain (e.g., Doran et al.~2013).

\section{Discussion and Conclusions}
\label{sec:discussion}

We find that the apparent contradiction between the properties (age/extinction) of young cluster populations in the near-IR and optical reported in the literature (with near-IR studies finding that clusters remain highly extincted for $\sim7$~Myr) is largely due to the aperture size and resolution of the studies. Each of the YMCs used in the current study have neighbouring sources and nebular emission that are bright in the near-IR, which significantly alters the integrated colours.  Since clusters are largely colourless in the near-IR after $\sim7$~Myr (e.g., Gazak et al.~2013), clusters that are blue quickly become redder as larger apertures are used (due to neighbouring contamination), ending up in the ``red cloud" in the near-IR colour space.  Hence, when large apertures are used, clusters that are young (i.e., have clear Br$\gamma$ associated with them) will appear red, and will be assigned high extinction values.

On the other hand, optical studies do not appear to be as heavily affected by resolution.  This is due to two effects.  The first is that the natural extent of cluster colours in the optical is larger, ranging $\sim1.5$~mag in the $U-B$ colour, relative to $\sim0.5$~mag in $J-H$ or $H-K$.  Secondly, contamination from nearby (older) sources is much stronger in the near-IR than in the optical (Fig.~\ref{fig:flux_vs_age}).  This contamination forces the measured integrated colours (of the cluster and surroundings) to the ``red cloud" of points.  This in turn leads to erroneous inferences about the extinction (and possibly age, although this is usually determined by the presence/absence of emission line flux) of these clusters.  Hence, unless high resolution ($\lesssim20$~pc) apertures can be employed, it is not possible to infer the age or extinction distributions of clusters with broadband photometry in the near-IR, potentially invalidating recent results based on larger apertures ($45-100$~pc - e.g., Grosb{\o}l \& Dottori~2012, 2013). In general, using aperture sizes closest to the size of the clusters (radii of $5-10$~pc) leads to the best results.

Contamination from nearby sources has a larger effect on near-IR than optical colours.  For young clusters, we found that using aperture sizes between 11 and 87~pc did not fundamentally alter the optical colours, and hence the estimated properties, of YMCs.  This is in agreement with previous photometric studies of clusters from the ground and HST, that found consistent results (e.g., Larsen~2002).  

By studying how the spectral energy distribution of clusters change as a function of aperture size, we found that larger apertures result in a significant near-IR flux excess in some young clusters.  This is caused by the contribution of neighbouring sources as well as the nebular  emission from nearby (often related) ionised gas which is stronger at redder wavelengths (e.g., Reines et al.~2010; Adamo et al.~2010a). Models that include nebular emission can explain some of this excess near-IR emission, however, for some clusters the models cannot account for the additional flux.  It is in these sources that nearby stellar contamination in the near-IR is dominant.

Finally, we note that increasing the aperture size significantly affects the amount of nebular line emission found for the clusters (see the bottom panel in Fig.~\ref{fig:images}).  This is due to the fact that young clusters have cleared out a large amount of the (ionised and neutral) gas/dust around them, leaving a shell surrounding the cluster, the size of which should depend on the age and mass of the cluster and well as the density of the surrounding ISM.  When larger apertures are used, more of the shell is included, leading to larger amounts of ionised gas in the aperture, and strong nebular emission.

 In a future work (Adamo et al.~in prep.) we will use near-IR integral field spectroscopy of many of the regions presented here, to quantify the amount of the ``red excess" that is due to ionised gas and stellar contamination.  Additionally, in Hollyhead et al. (in prep.), we will explore the effect of including emission lines (e.g., H$\alpha$) in the SED fitting, in particular in cases where the clusters have driven large ionised bubbles outside the size of the aperture.

\section*{Acknowledgments}

We thank Preben Grosb{\o}l and Horacio Dottori for insightful discussions and the referee for a careful reading of the manuscript and for helpful suggestions.  NB and MG are partially funded by a Royal Society University Research Fellowship.

\bsp
\label{lastpage}
\end{document}